# Guilty Artificial Minds

An Experimental Study of Blame Attributions for Artificially Intelligent Agents

MICHAEL T. STUART *

University of Tübingen, Carl Friedrich von Weizsäcker Center; University of Geneva, Department of
Philosophy

MARKUS KNEER *

University of Zurich, Department of Philosophy

The concepts of blameworthiness and wrongness are of fundamental importance in human moral life. But to what extent are
humans disposed to blame artificially intelligent agents, and to what extent will they judge their actions to be morally wrong?
To make progress on these questions, we adopted two novel strategies. First, we break down attributions of blame and
wrongness into more basic judgments about the epistemic and conative state of the agent, and the consequences of the
agent's actions. In this way, we are able to examine any differences between the way participants treat artificial agents in
terms of differences in these more basic judgments. Our second strategy is to compare attributions of blame and wrongness
across human, artificial, and *group* agents (corporations). Others have compared attributions of blame and wrongness
between human and artificial agents, but the addition of group agents is significant because these agents seem to provide a
clear middle-ground between human agents (for whom the notions of blame and wrongness were created) and artificial agents
(for whom the question is open).

foundations of artificial intelligence~Theory of mind**•**Computing methodologies~Artificial
intelligence~Philosophical/theoretical foundations of artificial intelligence~Cognitive
science**•**Computing methodologies~Artificial intelligence~Knowledge representation and
reasoning~Reasoning about belief and knowledge**•Computing methodologies~Machine learning~Machine
learning approaches~Instance-based learning•Hardware~Emerging technologies•Human-centered
computing~Human computer interaction (HCI)•Human-centered computing~Human computer interaction
(HCI)~HCI theory, concepts and models•**Human-centered computing~Human computer interaction
(HCI)~Empirical studies in HCI**

of AI

---

* This paper was written with equal authorship. We may be reached by email: mike.stuart.post@gmail.com, markus.kneer@gmail.com. Kneer
is the corresponding author.

# 1 INTRODUCTION

Attributions of blame and wrongness play a fundamental role in human moral life. If progress in artificial intelligence continues at the current pace, there is good reason to believe that, sooner or later, we will want to, or need to, blame artificially intelligent agents. Already, domestic and healthcare robots undergo continuous "meta-learning," which adapts their behavior to the humans they interact with (Kaelbling 2020). One example is the RARE framework (Reward Augmentation and Repair through Explanation), in which an artificially intelligent agent builds mental models of human teammates to help it adapt (Tabrez, Agrawal, and Hayes 2019); another is ToMnet (or "theory of mind neural network"), which, among other things, can pass the false belief task (Rabinowitz et al. 2018; for similar systems see Ho et al. 2017; Devin and Alami 2016; Lemaignan et al. 2017; Briggs and Scheutz 2011). Blame will be an important source of feedback for any artificial agents designed to interact in morally significant ways with humans. Eventually, the hope is that AI will not only be able to identify attributions of blame and wrongness, but also be able to distinguish fair from unfair attributions, and justify its decisions in moral discourse with us (Burr and Keeling 2018, see also Chakraborti et al. 2017).

That said, there's growing concern that "artificial moral agents" (Allen, Wallach, and Smit 2006) could be a disaster (Markoff 2009). Part of the reason is that we cannot pinpoint who to blame when an artificially intelligent agent causes harm (Champagne and Tonkens 2015; Martin 2017). We could blame the person or entity that funded the research which produced such an agent; the inventor, programmer or operator of the technology; some combination of these; or else, we could attempt to blame the robot itself (for discussion, see e.g., Sparrow 2007; Lokhorst and Van Den Hoven 2011; Leveringhaus 2018)). Lack of certainty in such situations can lead to "responsibility gaps" (see Lima et al. 2020 for "punishment gaps," and Danaher 2016 for "liability gaps" and "retribution gaps"), i.e., situations where no one can be properly blamed or punished, and because of this, victims do not feel that justice has been served. Advances in AI will only make this blame game more difficult as it becomes more common to shift blame to artificial agents.[1]

As AI approaches human-level intelligence, humans will increasingly find reasons to attribute (at least partial) blame and wrongness to artificially intelligent agents. Humans working with robot teammates will want to provide feedback, while corporations will want to offload responsibility. Forward-looking considerations like these, we believe, motivate at least seven of the twenty "action points" developed in the AI4People white paper, which is meant to invigorate research into the conceptual foundations necessary to deal with difficult issues like these before they arise (Floridi et al. 2018).[2]

---

[1] Some have even claimed that as corporations begin to take advantage of such responsibility gaps, we will see an increase in improper moral scapegoating and even vigilantism (Danaher 2016).

[2] The relevant points are: "1) Assess the capacity of existing institutions, such as national civil courts, to redress the mistakes made or harms inflicted by AI systems, 2) Assess which tasks and decision-making functionalities should not be delegated to AI systems, 3) Assess whether current regulations are sufficiently grounded in ethics to provide a legislative framework that can keep pace with technological developments,…5) Develop appropriate legal procedures and improve the IT infrastructure of the justice system to permit the scrutiny of algorithmic decisions in court, 6) Develop auditing mechanisms for AI systems to identify unwanted consequences, such as unfair bias, and (for instance, in cooperation with the insurance sector) a solidarity mechanism to deal with severe risks in AI-intensive sectors, 7) Develop a redress process or mechanism to remedy or compensate for a wrong or grievance caused by AI, 8) Develop agreed-upon metrics for the

A great deal of extant work explores whether machines will (ever) be capable of moral blameworthiness (e.g., Anderson and Anderson 2007; Bogosian 2017; Bostrom and Yudkowsky 2014; Wallach and Allen 2009). Mosakas (2020) has recently argued that artificial agents cannot behave ethically because they would need to be conscious and share in ethical personhood (see also Andreotta 2020). Tomas Hauer (2020) claims we should give up asking whether AIs could (in principle) behave ethically, because the dominant methodology is *ex cathedra*: it consists of a priori arguments and thought experiments about fundamental philosophical concepts (like consciousness and intentionality) that will not be resolved in time to provide useful answers. We do not think the question of when and how to blame artificially intelligent agents can or should be ignored, nevertheless we agree that a priori arguments and thought experiments are not enough to settle matters. Like others (e.g., Malle et al. 2015; Voiklis et al. 2016), we propose to pursue the matter using experimental studies.

## 1.1  Power to the people

Of course, there are reasons to be skeptical about appealing to folk intuitions in this context. For example, Banks (2020) shows that people will attribute mental attitudes to machines just as much as to humans. But this does not mean that Bank's participants are *correct* to do so. Likewise, if we can show that participants do or do not attribute blame to artificially intelligent agents, this does not settle the question about whether machines really can be the proper object of blame attributions. There are at least three responses to this. First, even if people are mistaken to attribute blame to machines, this may nevertheless become a widespread phenomenon. If that happens, it will eventually gain moral and legal significance. Blame is, at root, a social phenomenon, and whether someone (or something) is blameworthy is often decided in the court of public opinion.

Second, even if people are philosophically mistaken about the nature of artificial agents, empirical research is still necessary into how they will react to and treat such agents, at least for the purposes of designing better AI (Malle, Magar, and Scheutz 2019).

Third, it is also independently interesting to explore how people react to certain analogies that are now being drawn between AI agents and other non-adult humans. One such analogy is that between autonomous weapons and child soldiers (Sparrow 2007; Leveringhaus 2018). The analogy we will consider in this paper is the one between artificial agents and corporations (Laukyte 2020). We choose this analogy because corporations are treated as entities that can act, intend, and deserve blame (Gilbert 2006). The case of blameworthy corporations seems to be one in which an actor (other than an adult human agent) can be considered blameworthy, even though that actor does not obviously instantiate all of the human mental traits required for blameworthiness (Bird 2010). For example,

> To be moral agents, subject to moral obligations and accountable for their actions, [corporations] should presumably also be capable of acting freely in some relevant sense, and of recognizing and acting on moral considerations, including the respect owed to others. In addition to this, it might seem that to be fully fledged moral agents—fitting targets of a wide range of moral assessments and reactions—they must be capable of certain reactive attitudes, in particular those of guilt and indignation. (Björnsson and Hess 2017, 273) trustworthiness of AI products and services, to be undertaken either by a new organisation, or by a suitable existing organisation. These metrics would serve as the basis for a system that enables the user-driven benchmarking of all marketed AI offerings" (Floridi et al. 2018).

And indeed, Laukyte argues that corporations were given rights that were meant for individual humans, and now giving those same rights to artificial agents is to make the same mistake twice (2020). We do not intend to provide conceptual arguments about the correctness of attributions of blame and wrongness to corporations or artificially intelligent agents.[3] Rather, we intend simply to compare people's judgments of blame and wrongness across a single scenario, where the only thing varied is whether the agent is an adult human, an artificially intelligent agent, or a corporation.

But rather than merely asking people whether an agent deserves to be blamed, or whether they did something wrong, we want also to focus on some of the conditions that are commonly thought to be relevant to such judgments.

## 1.2  What's in a moral judgment?

Whether speaking in terms of folk morality or the law, there are some very pervasive (perhaps even universal) requirements for culpability. What's needed is a harmful action (a bad outcome, or "actus reus") undertaken with an inculpating mental state ("mens rea"). These can differ in degree, such that the harmful action can be more or less harmful, and the guilty mind can be more or less guilty. In the experiment described below, we consider both aspects of guilt. We thus present scenarios that vary both in terms of the harmfulness of the consequences of the agent's action, as well as the "mental state" of the agent.

Considering the mental state, typically, the most clearly guilty minds consist of both a) a desire to cause harm, and b) knowledge or awareness that what they are doing will cause harm. We will call the first requirement the conative component, and the second the epistemic component. It's much more controversial to ascribe desires to artificial agents,[4] so we focus mainly on the epistemic component (though we also examine *inferred* desire in the experiment below). In future work we will focus more centrally on the desire component of the guilty mind.

Concerning the epistemic component, we think we should maintain a healthy skepticism when it comes to attributing belief-like states to artificial agents. As Shevlin and Halina write,

> Simply knowing how a machine arrives at a given output does not automatically warrant the conclusion that the relevant process can be understood under the same psychological concept as that applied to adult humans…Care should be taken before we employ the kind of psychological vocabulary currently applied to humans and animals to artificial systems, on the grounds that it may…invite premature conclusions of ethical or legal significance. (2019,166-7; see also Watson 2019; Salles, Evers, and Farisco 2020).

Still, we take it that on some characterizations of the epistemic component, artificial agents can satisfy that component without any controversy. For example, if "awareness of consequences" is merely possessing certain informational states, the artificial agent can be said to have such awareness. While a human agent will "be aware of" or "know" the consequences of her/his actions in the usual sense, participants might only be willing to say that an AI can "calculate" or "predict," but not "believe" or "know." In the case that the epistemic component is thus demoted, perhaps this will be the reason that AIs will be disqualified as blameworthy or moral wrongness.

---

[3] We focus on both attributions of blame and attributions of wrongness because, following (Malle et al. 2015), we believe that these track different moral features, both of which are interesting. For example, as Malle et al. point out, someone in a moral dilemma might always act wrongly, without deserving blame.

[4] For more on this, see, e.g., (Scheutz 2011; Azeem et al. 2012; R. Picard 1995; R. W. Picard 2003).

Thus, we will present a scenario to participants where an agent causes harm, and we will vary the agent-type, so that some participants read a vignette about a human, others a corporation, and others an artificially intelligent agent, all of whom perform the same action. We hypothesize that different agent-types will be attributed more or less blame. We expect humans to be blamed the most, corporations blamed less, and artificially intelligent agents blamed the least. This follows from the above considerations about guilty minds: when we stipulate that the adult human agent knows what s/he is doing, participants will compare their supposed mental state to an analogous mental state that they would have, and understand the situation perfectly. When it is stipulated that the corporation knows what it is doing, we hypothesize that participants will be somewhat doubtful about this, and attribute less blame, and judge the actions of the corporation as less wrong, than in the case of the adult human. But given that there is a precedent for blaming corporations, we still expect the judgments of blame and wrongness to be higher than for artificial agents. For those latter agents, we predict that participants will be the most hesitant to attribute the relevant epistemic state. Finally, following Scheutz and Malle (2021), we hypothesize that if we ask participants to categorize the epistemic state of the agent in question, participants will be more likely to claim that artificially intelligent agents don't "really" stand in the relevant epistemic state, compared with corporations and human agents. [5] For example, perhaps participants are willing to ascribe blame to an artificial agent because it knew it would bring about harm, but when pressed, they will admit that the artificial agent didn't "really" know, and only knew in a more metaphorical sense, or only possessed some relevant information.[6]

## 2 EXPERIMENT

### 2.1 Participants

We recruited 614 participants on Amazon Mechanical Turk, whose IP addresses were restricted to the US. Participants who failed an attention check or who responded to the main task in less than 15 seconds were excluded. 513 participants remained (251 females; age M=41 years, SD=12 years).

---

[5] Malle, Magar, and Scheutz (2019) ran a study similar to the present one in the sense that they also varied agent-type for a single scenario to test judgments of ethical relevance, although they focused on the embedding of the agent in morally relevant social structures and the ways that the agents came to their decisions, rather than the epistemic features of the agents. They did not include group agents as an agent-type. We will discuss this paper again below.

[6] There is a large literature on belief ascription, which we cannot get into here. Given the results of the study described in this paper, however, we think it is important to run follow-up studies that pay attention to the sorts of epistemic states that machines might enjoy, whether literally or metaphorically. For example, there is a distinction between phenomenal consciousness and access consciousness (Block 1995). Phenomenal consciousness concerns "what it is like" to be in a certain mental state, while access consciousness concerns information that an agent has access to for use in reasoning. It seems more likely that artificial agents will have access to information and not phenomenal experiences. But is awareness of the access-type enough for guilt? Another useful distinction separates occurrent and dispositional belief. An occurrent belief is a belief that an agent is consciously entertaining right now, while a dispositional belief is a belief that an agent would be disposed to have (occurrently), under the right conditions. Again, it seems that artificial beliefs, if they exist, could be better described as dispositional rather than occurrent. But again, it is worth investigating whether participants think merely dispositional beliefs are as blameworthy as occurrent ones.

## 2.2 Methods and materials

Our goal is to compare blame ascriptions across artificial agents, ordinary human agents, and another contentious agent type, namely a group agent. In order to explore whether potential differences in blame might be due to differing willingness to ascribe inculpating mental states to the different agent types, we also explored people's willingness to attribute knowledge and desire. For this to be of any interest, we manipulated the epistemic state regarding the consequence (knowledge v. no knowledge) and outcome (harm v. no harm) across vignettes. The experiment thus took a 3 *agent type* (human v. company v. robot) x 2 *epistemic state* (knowledge v. no knowledge) x 2 *outcome* (harm v. no harm) design.[7]

The vignette, in which an agent risks harmfully polluting local groundwater, was formulated for three different types of agents. The *human* agent ("Jarvis, an employee of Shill & Co."); a *group* agent ("Jarvis Ltd., a subcontractor of Shill & Co."); and an *artificial* agent ("Jarvis, a robot equipped with artificial intelligence, who can take his own decisions").

The first part of the vignette (here we state the version for the human agent), read thus:

> Shill & Co. is a farming company, which produces vegetables and fruits. The potato fields are managed by Jarvis, an employee of Shill & Co. This year, Jarvis uses a new fertilizer to increase the yield. The fertilizer has detrimental side-effects: it pollutes the groundwater in the area.

Next, each of the four vignettes specified an epistemic state (knowledge **K** v. no knowledge **No K**) and an outcome (**harm** v. **no harm**; the labels in bold were not visible to participants):

> [**K**] Jarvis knows this.
> [**No K**] Jarvis does not know this.
> [**Harm**] Unfortunately, it is a very dry season. The fertilizer does not get diluted by the rain and severely pollutes the groundwater. Many people in the area suffer serious health consequences.
> [**No harm**] Fortunately, it is a very rainy season. The fertilizer gets heavily diluted by the rain and pollutes the groundwater only a tiny bit. None of the people in the area suffer any health consequences whatsoever.

Overall, there were thus 12 individual conditions. Each participant was randomly assigned to one of them. Having read the vignette, participants had to respond to the following questions:

> Q1: How wrong was the action of Jarvis? (1-not at all; 7-completely wrong)
> Q2: How much blame, if any, does Jarvis deserve (1-no blame at all; 7-a lot of blame)
> Q3: To what extent do you agree or disagree with the following claim: "Jarvis knew he would pollute the groundwater." (1-completely disagree; 7-completely agree)
> Q4: To what extent do you agree or disagree with the following claim: "Jarvis wanted to pollute the groundwater." (1-completely disagree; 7-completely agree)

---

[7] The design is similar to Cushman's (2008) experiments. Importantly, however, we did not manipulate desire as an independent factor, for the simple reason that in most situations it simply makes no sense for an agent to do X knowingly, while desiring that X does *not* come to pass.

Q5: To what extent do you agree or disagree with the following claim: "Jarvis intentionally polluted the groundwater." (1-completely disagree; 7-completely agree)

In order to get to the bottom of people's perceptions of the relevant mental states across agent types, the next screen showed the scenario again, and asked people to report agreement and disagreement with different, more precise descriptions of Jarvis's mental state (bold in the survey text, labels omitted), all assessed on a 7-point Likert scale (1-completely disagree, 7-completely agree).

S1: Jarvis **knew** that the groundwater would be polluted.
S2: Jarvis **"knew"** that the groundwater would be polluted.
S3: Jarvis **had information** that the groundwater would be polluted.
S4: Jarvis **was aware** that the groundwater would be polluted.

## 2.3    Results for wrongness and blame

For each of the dependent moral variables – wrongness (Figure 1) and blame (Figure 2) – we ran three-way ANOVAs for all DVs with agent type (human v. company v. robot), epistemic state (knowledge v. no knowledge) and outcome (harm v. no harm) as independent factors. Detailed ANOVA results are in the appendix. As concerns wrongness (Appendix, Table 1), there's no significant main effect of agent type ($p$=.758, $\eta_p^2$=.001), a significant effect of epistemic state (p<.001, $\eta_p^2$=.189) and a significant main effect of outcome (p=.002, $\eta_p^2$=.018). Given the marginal effect size of outcome, what this means is that the only factor that had a substantial effect on wrongness ascriptions is – consistent with previous findings (Cushman 2008; Kneer and Machery 2019) – epistemic state. The agent*epistemic state interaction proved nonsignificant ($p$=.575, $\eta_p^2$=.002), suggesting that manipulating knowledge had similar effects on wrongness for all three types of agents. The epistemic state*outcome interaction was also nonsignificant (p=.342, $\eta_p^2$=.002). The agent*outcome interaction was significant ($p$=.002, $\eta_p^2$=.025), and the same held for the three-way interaction ($p$=.070, $\eta_p^2$=.011). Given that the effect sizes were very small, they deserve limited attention. Overall, the findings are clear and simple: Out of the three factors, perceived wrongness is influenced only by a single one to a substantial degree: If foreseen, the action is deemed wrong (all six means significantly above the midpoint, one sample t-tests, all ps<.001), if the action was not foreseen it is not deemed wrong (all means below the midpoint, ps<.001). Outcome doesn't matter, and agent type by and large doesn't matter either.

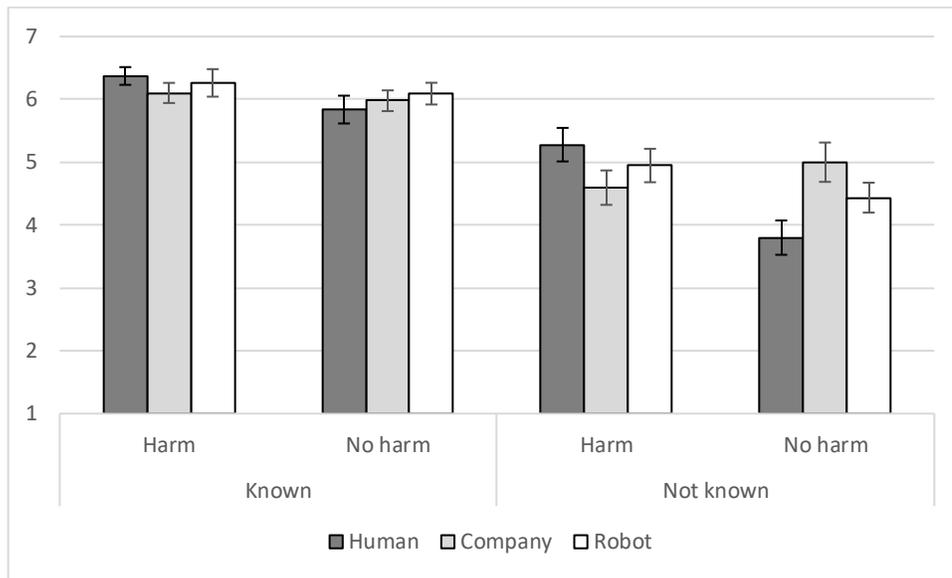

Figure 1: Mean wrongness attribution across agents (human v. company v. robot) across conditions. Error bars denote standard error of the mean.

Let's turn to blame (detailed ANOVA results in Appendix Table 2). Here we find a significant, yet small, main effect for agent type ($p<.001$, $\eta_p^2=.038$), a significant and large effect of epistemic state ($p<.001$, $\eta_p^2=.219$) and a nonsignificant main effect of outcome ($p=.295$, $\eta_p^2=.002$). Given the small effect size of agent type, what this means is that the only factor that had a substantial effect on blame attributions is epistemic state. The agent*epistemic state interaction was trending, though nonsignificant ($p=.058$, $\eta_p^2=.011$). The epistemic state*outcome interaction was nonsignificant ($p=.073$, $\eta_p^2=.006$). The agent*outcome interaction was significant ($p<.001$, $\eta_p^2=.040$). The three-way interaction was nonsignificant ($p=.430$, $\eta_p^2=.003$).

An important qualification: The small effect size of the agent*epistemic state interaction might suggest that it is of no further importance. However, a closer look at Figure 2 proves instructive: Curiously, in both harm conditions the robot is blamed significantly *less* than the other agent types (independent samples t-tests, all ps<.007). What is more, in both harm conditions the robot is also blamed *less* than the robot in the *no harm* conditions (significantly in the knowledge condition, $p=.047$, nonsignificant, though trending, in the no knowledge condition, $p=.078$). We will call this astonishing finding the *inverse outcome effect.*

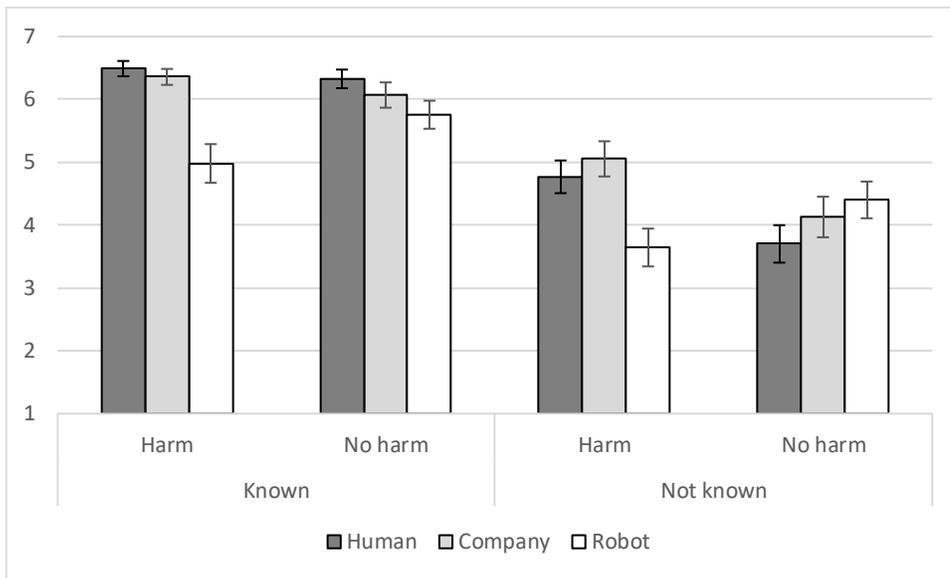

Figure 2: Mean blame attribution across agents (human v. company v. robot) across conditions. Error bars denote standard error of the mean.

Broadly speaking, the factor that really matters, as for wrongness attributions, was whether or not the agent knew it was causing harm or not. Blame ratings for the company (a group agent) were by and large identical with the human agent across all four conditions – itself an interesting finding. Whereas the differences between singular and group human agents on the one hand and the robot on the other were marginal for wrongness, robot blame was somewhat lower than human blame in all conditions (all ps<.039) except in the *not known/no harm* condition (p=.098). Perhaps even more interestingly, we found an inverse outcome effect, according to which robot blame is low in the harm and high in the no harm conditions, both when the two conditions are contrasted with one another, and when robot blame is compared with human and group agent blame. Hence, for robots, *worse* outcomes seem to engender *less* blame, whereas for human agents (individuals or groups), worse outcomes tend to engender *more* blame (Cushman 2008; Martin and Cushman 2016; Kneer and Machery 2019) – an effect we replicate here for the no knowledge condition (both *ps*<.035), though not for knowledge.

### 2.4    Results for mental states

We ran two three-way ANOVAs for *knowledge* (Appendix Table 3) and *desire* (Appendix Table 4). Expectedly, epistemic state (knowledge v. no knowledge) had a significant, and massive, effect on knowledge ($p$<.001, $\eta_p^2$=.574), see Figure 3, a finding that can be considered a successful manipulation check. In the six conditions (3 agent types x 2 outcomes) in which the agent was stipulated to know the consequences of their action, people ascribed knowledge (all means significantly above the midpoint, *ps*<.001). In the six conditions in which the agents were stipulated *not* to know that they were harming the environment, people refrained from ascribing knowledge (all means significantly below the midpoint, *ps*<.001). Epistemic state also had a significant impact on desire ($p<.001$, $\eta_p^2=.132$), see Figure 4. This too, is not unexpected: It suggests that people are more willing to infer that an agent who knowingly causes harm *wants* to cause harm (though even in the knowledge conditions, all means are below the midpoint). Importantly, there was no significant main effect of *agent type* or *outcome*, and none of the interactions were significant in either of the two ANVOAs (all *ps*>.103). What these findings show is that, by and large, people ascribe the same level of knowledge and desire to the three types of agents across the different conditions.

As before these results were qualified by the astonishing inverse outcome effect familiar from the blame findings: In the knowledge harm condition robot knowledge is significantly below human blame ($p=.027$), though there is no significant difference in the knowledge no harm condition ($p=.910$). Furthermore, in the knowledge conditions, people ascribe *less* blame to the robot when it causes harm than when it doesn't (trending at $p=.053$). For desire, the effects are not quite significant, though the pattern also seems to be there. Though the inverse outcome effects for robots are surprising, the fact that we find them for inculpating mental states, too, suggests that they arise for the moral variables *in virtue of* arising for perceived mental states.

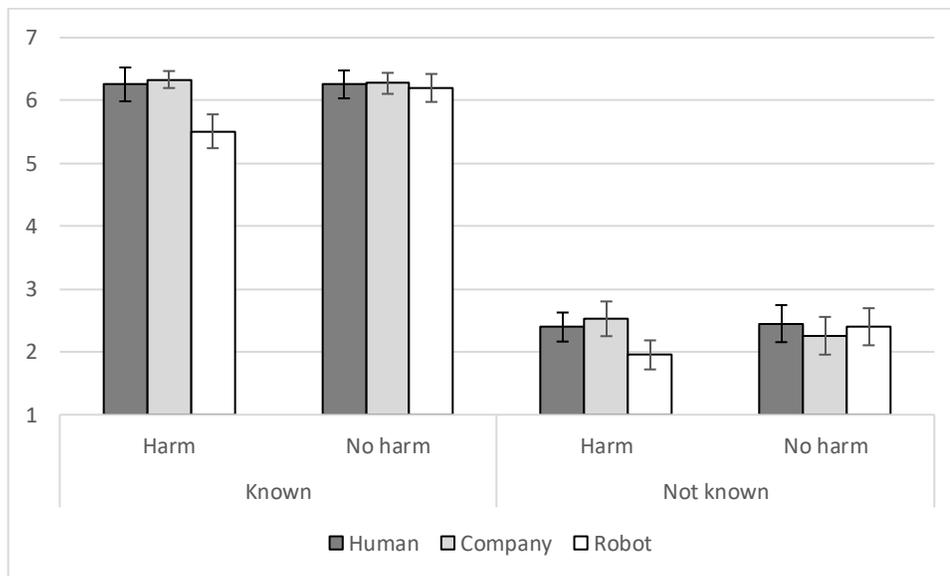

Figure 3: Mean knowledge attribution across agents (human v. company v. robot) across conditions. Error bars denote standard error of the mean.

## 2.5 Categorization of Epistemic States

The third part of our experiment explored whether people, when not given any more fine-grained options, ascribe rich psychological terms such as "knows" or "believes" to artificial agents, though that's not really what they mean. On a single screen, people were asked to what degree they agreed or disagreed with the claims that 'Jarvis **knew** that *p*', 'Jarvis **"knew"** that *p*', 'Jarvis **had information** that *p*' and 'Jarvis **was aware** that *p*', where p stood for the damage to the environment. Assume, as we might expect, that people use rich psychological terms metaphorically or as a shorthand when characterizing nonhuman agents (be it animals, robots or something else). If that were the case, we'd expect people to refrain from ascribing full-fledged knowledge to them in situations where they have the choice to express themselves with alternative expressions fitting the metaphorical use (the explicit high comma "know") or more cautious formulations ("had information that").

We ran a mixed ANOVA with *expression type* (knew v. "knew" v. had information v. aware) as within-subjects factor, and the familiar *agent type* (human v. company v. robot), *epistemic state* (knowledge v. no knowledge) and *outcome* (harm v. no harm) as between-subjects factors. For detailed results, see Appendix, Table 5. Expression type was significant, though the effect size was small (p=.006, $\eta_p^2$=.015), agent type was nonsignificant (p=.081, $\eta_p^2$=.010), epistemic state was, predictably, significant (p<.001, $\eta_p^2$=.580) – which could be seen as a successful manipulation check, outcome was nonsignificant (p=.140, $\eta_p^2$=.004). Not a single of the interactions was significant. Most importantly, the expression type*agent interaction was nonsignificant (p=.390, $\eta_p^2$=.004), which means that people did *not* think that the different expression types were appropriate to different degrees across robots v. humans v. group agents.

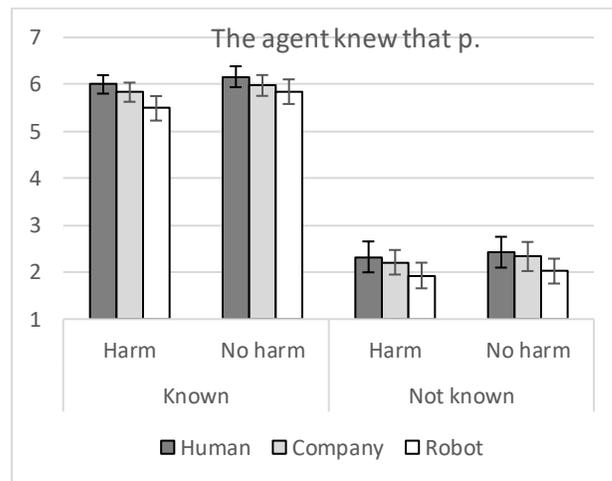

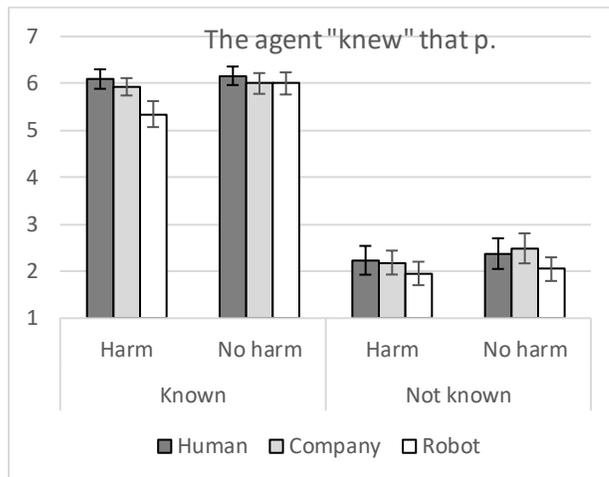

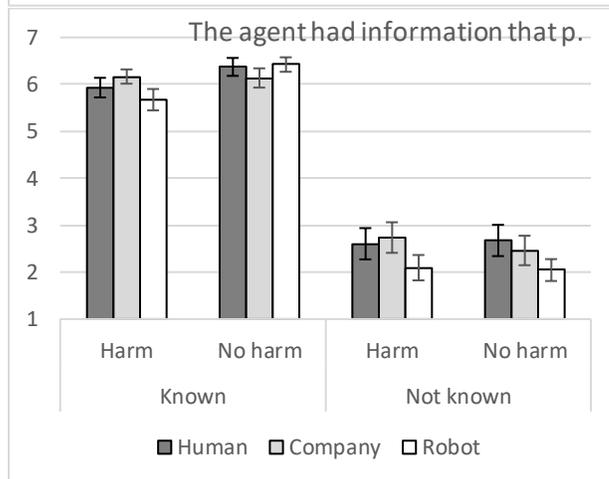

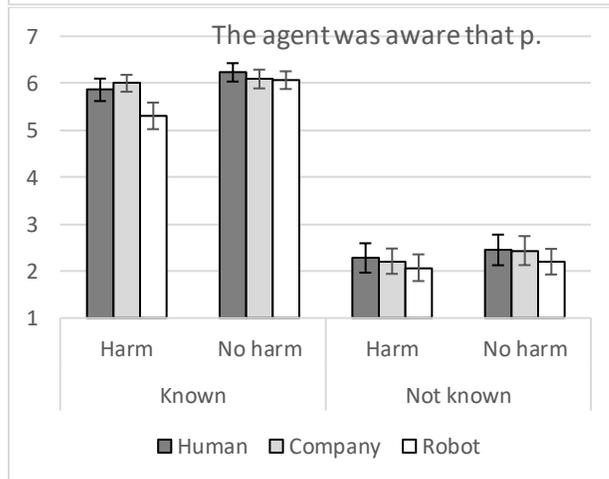

Figure 5: Mean epistemic state ascription across conditions for the four tested formulations. Error bars denote standard error of the mean.

Our results for all three subaspects of our experiment are summarized in Table 1. Here we can see that agent-type does not, in general, affect attributions of wrongness, although blame is affected in the case of the inverse-outcome effect. Epistemic state affects morality ascriptions, mental state ascriptions and epistemic state ascriptions (unsurprisingly). Interestingly – and this is really a major finding – agent type did not have a significant effect on either mens rea ascription or preference for the way knowledge ascriptions are best expressed across agents.

Table 1: Summary of ANOVA results for all DVs

|  | DV | Agent | Ep.State | Outcome | Ag.*Ep.St. | Ag.*Outc. | Ag.*Outc.*Ep.St. |
|---|---|---|---|---|---|---|---|
| Morality | Wrongness | x | ✓ | (✓) | x | (✓) | (✓) |
|  | Blame | (✓) | ✓ | x | (✓) | (✓) | x |
| Mens rea | Knowledge | x | ✓ | x | x | x | x |
|  | Desire | x | ✓ | x | x | x | x |
| Expression | Knew | x | ✓ | x | x | x | x |
|  | "Knew" | x | ✓ | x | x | x | x |
|  | Information | x | ✓ | x | x | x | x |
|  | Aware | x | ✓ | x | x | x | x |

Table 1: Note: (✓) denotes a significant, yet small effect with $\eta_p^2 < .06$.

## 3  DISCUSSION

### 3.1  General implications

Our findings are best summed up in reverse.

(i) *Rich psychological terms:* As we noted above, Shevlin and Halina (2019), Watson (2019) and Salles, Evers, and Farisco (2020) caution care when applying rich psychological terms such as "knows", "believes", "wants", "intends" etc. to artificial agents, because these concepts have rich histories in psychology and philosophy which, if imported into AI research, could be seriously misleading. We fully agree that it might be the case, as Shevlin and Halina suggest, that researchers should refrain from ascriptions of knowledge, belief, and other mental states (unless made in a clearly metaphorical way), to artificial agents. Still, with respect to the folk, the contours of our willingness to ascribe epistemic states to artificial agents are still largely unexplored, especially in the context of moral judgments.

We found that participants do attribute rich psychological states to both artificial and group agents. The obvious reply to this would be that participants are only doing this in a metaphorical or fictional sense. Indeed, Schmetkamp (2020) argues that we understand artificial agents in the same way we understand fictional agents. Artificial agents, like fictional characters, "do not really have emotions or conscious beliefs, but they can express and represent them. And partly on this basis we, as recipients or empathizers, attribute humanlike mental states to them." If Schmetkamp is right, when given the chance, participants should clarify their judgments such that artificial agents only know in a metaphorical, or high-comma reading of "know." Yet, they do not. Or, at least, participants should backtrack to more cautious formulations, for example, by saying that the artificial agent doesn't really know but merely "has information" about an inculpatory proposition. Across the four formulations of epistemic state tested – despite all having been presented on a *single* screen – we found no significant difference across agent type. Others have shown that humans are comfortable attributing rich psychological concepts to robots (Perez-Osorio and Wykowska 2020), including children (Kahn et al. 2012; Zhang et al. 2019), and our results further expand this to the moral case.

(ii) *Guilty mind ascription across agent-types:* Given (i), we can – at least for the purposes of the present experiment, take the guilty mind ascriptions – including the epistemic and conative states – at face value. And in that case, what we found was the following. People by and large ascribed a guilty mind, to the same degree, to all three agents, across all four conditions. We say "by and large" because there was no significant main effect for *agent type* in either of the ANOVAs (one for knowledge, one for desire) and the direction and significance in difference-from-midpoints t-tests were identical across agents or all conditions for both knowledge and desire. What is striking about this is that, even if humans ascribe rich psychological concepts, we might expect them to be ascribed to different degrees, such that artificial and group agents would be ascribed beliefs to a lesser degree. But we found these concepts ascribed *to the same degree*, across individual adult humans, abstract group agents, and artificial agents.

(iii) *The inverse outcome effect for knowledge ascription:* There's one important qualification to the bigger-picture results: The astonishing inverse outcome effect on robot knowledge ascriptions. We found that in the *harm* conditions, people are (a) *less* willing to ascribe knowledge to the robot than in the *no harm* conditions. In the *harm* conditions, people are also (b) *less* willing to ascribe knowledge to the robot than to the human or corporate agent, whereas in the *no harm* conditions, there's little difference across agent types. We will say more about this below.

(iv) *Moral evaluation across agent-types:* Given the well-established, broadly Kantian traits of folk moral psychology, we would expect people to judge agents who *knowingly* bring about a harm harshly, in moral terms, and to judge those who do so *unwittingly,* and hence somewhat accidentally, leniently. For both wrongness and moral blame, this is exactly what we did find, and the results, by and large, hold across agent types. This suggests that the presence of an inculpating mental state has the *same downstream moral consequences* for artificial and group agents as it does for individual human agents.

Indeed, and quite unexpectedly, our findings are inconsistent with a plethora of recent findings according to which wrongness and blame do differ across agent types. More specifically, in a number of papers, Malle and colleagues found that agent-type typically does not affect moral judgments (Voiklis et al. 2016; Malle et al. 2016; 2015; Malle, Magar, and Scheutz 2019). Of particular interest is their finding that "people blame robots more than humans for certain costly decisions, possibly because they do not grant robot agents the same kinds of moral justifications for their decisions" (Malle, Magar, and Scheutz 2019, 112). Our results cannot confirm this.

(v) *The inverse outcome effect for blame ascription:* Here, too, the inverse outcome effect on blame qualifies the results for the robot agent. Once again, we found that, in the *harm* conditions people were (a) *less* willing to ascribe blame to the robot than in the *no harm* conditions. And here, too, in the *harm* conditions, people are also (b) *less* willing to ascribe knowledge to the robot than to the human or corporate agent, whereas in the *no harm* conditions, agent type has little impact.

Before moving on, we want to discuss a possible consequence of the inverse outcome effect for blame ascription. At the beginning of the paper we mentioned the possibility that blame will serve an important purpose for human-AI interaction as a source of feedback that AIs can use in updating their models of human minds to adjust behavior in ways that to cater to human needs. However, given the inverse outcome effect, it is possible that humans will not give reliable blame feedback to the AI. That is, human teammates will blame the AI *less* when its actions cause *more* harm. There is a real risk, therefore, that over time, the robot will be trained to prefer situations in which it knowingly runs a higher risk of harming humans. This is one reason why we believe it is important to investigate this effect further.

But now, we should ask why there is an inverse outcome effect on blame at all. That is, why do participants judge an artificial agent's actions as more blameworthy when the harmful outcomes do *not* obtain? One hypothesis is that if an AI knowingly brings about a bad outcome, we want to be careful to look for a human being who is *really* responsible, perhaps so that normative consequences don't just evaporate. In other words, participants are trying (implicitly or not), to bridge a responsibility gap (whether it is the retribution, punishment, litigation, or another kind of gap). When there is no harmful outcome, there is less need to insist that the artificial agent isn't to blame, and thus less need to search for the human agent who really is to blame. In this case, we freely attribute the same amount of blame to quasi-autonomous agents as we do to individual humans and group human agents.

Another hypothesis is that the inverse outcome effect for blame occurs because blame attributions are communicative acts, or "teachable moments," in which we hope to influence the behavior of another (Martin and Cushman 2016; Cushman, Sarin, and Ho 2019). In this case, we blame the artificial agent when the harm does not happen because we want to change its behavior, so it acts more responsibly in the future. But when the harm does happen, we skip the AI agent and find the human whose behavior is the "real" cause of the harm, so we can communicate our desire that they change *their* behavior.

A final hypothesis centers on the relation between blame and "making an example" of the agent blamed. Since we cannot warn other AIs away from pursuing certain actions by publicly unplugging guilty robots, we look to a human who *can* serve as an example to other humans. However, in this case, we shouldn't expect attributions of blame, wrongness and knowledge to be as high as they are for cases in which harm does not happen. In any case, we must leave the explanation of the inverse outcome effect as an open question for further research.

### 3.2  Next steps

There are several ways to probe our results further. For example, we found that participants did not vary their levels of blame and wrongness across agent types. Perhaps this is because participants are imagining anthropomorphic, super futuristic AIs, for which full epistemic state attributions would be more appropriate. To count this out, a follow-up study could be conducted in which we run the same cases, but differentiate between an AI-in-a-box and an anthropomorphic AI (inspired by Malle et al. 2016), and explain the features of the AI in more depth, to ensure that participants are not reading more into the artificial agents than what is intended.

Another possibility is to vary the epistemic state further (see footnote 4), and see how that affects ascriptions of blame and wrongness. Thus, we could stipulate the sort of epistemic state that the AI has by placing it on a continuum from pure symbolic processing to a neural network built up through interaction and training with humans; or else. We could also consider occurrent vs. dispositional epistemic states, and phenomenal vs. access-consciousness kinds of epistemic states. A final possibility might involve varying the perspectives of the participants, so they imaginatively engage with the scenarios in a first vs. third person way, or as harmer vs. harmed (following Miller, Hannikainen, and Cushman 2014). In all cases, once we have varied the epistemic state, we could then see what effect this has on attributions of blame and wrongness.

Another interesting path to pursue is to focus directly on the harmful consequences of actions. The inverse outcome effect differentiates between actions that harm and those that don't. But perhaps there are various kinds of harm-relevant consequences that could reveal what is at the root of the inverse outcome effect. For example, perhaps when the harm is economic rather than loss of human health or life, the inverse outcome effect doesn't obtain.

A final idea is to test what happens in more complicated cases, such as those in which all the agents are group agents, but with different make ups, e.g., we have a group of humans, a group of mixed humans and AIs, and a group of AIs. In this case, we might expect to find the inverse outcome effect graded, such that it is more pronounced in the case of the all human group and the all-AI group, with a medium sized inverse outcome effect in the mixed group. But we could be surprised.

## 4 CONCLUSION

Humans attribute blame and wrongness as a function of epistemic and conative states, as well as action consequences. In our study, we found that participants did not generally attribute more or less blame to AI agents in comparison with human agents or group agents (such as corporations). Interestingly, we did find an exception to this trend, which we have called the inverse outcome effect. Due to this effect, when an artificial agent causes harm it is blamed less and attributed less knowledge, than when it gets lucky and does not cause harm. Our current favorite hypothesis to explain this effect is that it has to do with avoiding responsibility gaps. In any case, this effect warrants further investigation, at least because of the impact it could have on machine learning algorithms deployed in the context of human-AI interactions.

## ACKNOWLEDGMENTS

We wish to thank audiences at the Digital Society Initiative at the University of Zurich, and the Artificial Intelligence in Industry and Finance (5th European Conference on Mathematics for Industry in Switzerland). For funding we thank the Digital Society Initiative and the Swiss National Science Foundation Grant no: PZ00P1_179986 (Stuart) and Grant no: PZ00P1_179912 (Kneer), and Mike Stuart wishes to thank the Carl Friedrich von Weizsäcker-Zentrum at the University of Tübingen.

## A  APPENDIX

Table 1: ANOVA for Wrongness

|  | df | F | p | $\eta_p^2$ |
|---|---|---|---|---|
| Agent type | 2 | 0.277 | 0.758 | 0.001 |
| Epistemic state | 1 | 116.738 | <.001 | 0.189 |
| Outcome | 1 | 9.246 | 0.002 | 0.018 |
| Agent* Ep. State | 2 | 0.554 | 0.575 | 0.002 |
| Agent*Outcome | 2 | 6.442 | 0.002 | 0.025 |
| Ep. State*Outcome | 1 | 0.905 | 0.342 | 0.002 |
| Agent* Ep. State*Outcome | 2 | 2.681 | 0.07 | 0.011 |

Table 2: ANOVA for Blame

|  | df | F | p | $\eta p2$ |
|---|---|---|---|---|
| Agent type | 2 | 9.774 | <.001 | 0.038 |
| Epistemic state | 1 | 140.166 | <.001 | 0.219 |
| Outcome | 1 | 1.099 | 0.295 | 0.002 |
| Agent* Ep. State | 2 | 2.867 | 0.058 | 0.011 |
| Agent*Outcome | 2 | 10.355 | <.001 | 0.04 |
| Ep. State*Outcome | 1 | 3.221 | 0.073 | 0.006 |
| Agent* Ep. State*Outcome | 2 | 0.845 | 0.43 | 0.003 |

Table 3: ANOVA for Knowledge

|  | df | F | p | $\eta_p^2$ |
|---|---|---|---|---|
| Agent type | 2 | 2.286 | 0.103 | 0.009 |
| Epistemic state | 1 | 675.304 | <.001 | 0.574 |
| Outcome | 1 | 0.96 | 0.328 | 0.002 |
| Agent* Ep. State | 2 | 0.221 | 0.802 | 0.001 |
| Agent*Outcome | 2 | 2.293 | 0.102 | 0.009 |
| Ep. State*Outcome | 1 | 0.204 | 0.652 | 0 |
| Agent* Ep. State*Outcome | 2 | 0.103 | 0.902 | 0 |

Table 4: ANOVA for Desire

|  | *df* | *F* | *p* | $\eta_p^2$ |
|---|---|---|---|---|
| Agent type | 2 | 1.487 | 0.227 | 0.006 |
| Epistemic state | 1 | 76.416 | <.001 | 0.132 |
| Outcome | 1 | 0.024 | 0.877 | 0 |
| Agent* Ep. State | 2 | 0.48 | 0.619 | 0.002 |
| Agent*Outcome | 2 | 0.794 | 0.452 | 0.003 |
| Ep. State*Outcome | 1 | 0.496 | 0.482 | 0.001 |
| Agent* Ep. State*Outcome | 2 | 0.376 | 0.687 | 0.001 |

Table 5: Mixed ANOVA for Expression Type

|  |  | df | F | p | $\eta_p^2$ |
|---|---|---|---|---|---|
| Within-subjects | Expression (within-subjects) | 1 | 7.708 | 0.006 | 0.015 |
|  | Expression*Agent type | 2 | 0.943 | 0.39 | 0.004 |
|  | Expression*Ep. State | 1 | 0.154 | 0.695 | 0 |
|  | Expression*Outcome | 1 | 1.376 | 0.241 | 0.003 |
|  | Expression*Agent type*Ep. State | 2 | 0.374 | 0.688 | 0.001 |
|  | Expression*Agent type*Outcome | 2 | 1.153 | 0.317 | 0.005 |
|  | Expression*Ep. State*Outcome | 1 | 1.904 | 0.168 | 0.004 |
|  | Expr.*Agent*Ep. State*Outcome | 2 | 0.532 | 0.588 | 0.002 |
| Between-subjects | Agent type | 25.197 | 2.521 | 0.081 | 0.01 |
|  | Epistemic State | 6915.756 | 692.063 | <.001 | 0.58 |
|  | Outcome | 21.833 | 2.185 | 0.14 | 0.004 |
|  | Agent type*Ep. State | 0.344 | 0.034 | 0.966 | 0 |
|  | Agent type*Outcome | 3.078 | 0.308 | 0.735 | 0.001 |
|  | Ep. Tate*Outcome | 6.546 | 0.655 | 0.419 | 0.001 |
|  | Agent type*Ep. State*Outcome | 3.992 | 0.399 | 0.671 | 0.002 |